\documentclass[english]{css}

\usepackage{graphicx}
\usepackage{amsmath,amssymb}
\usepackage{booktabs}
\usepackage{pgfplots}
\pgfplotsset{compat=1.18}
\usepackage[hyphens]{url}

\begin{document}

\title{Bug Bounty Programs in the AI Era: How Does Platform Anti-Spam Friction Affect Vendor Profitability?}

\affiliate{WU}{Waseda University, Sako Laboratory}
\affiliate{NYX}{Nyx Foundation}

\author{Haruki Oyama}{WU}[harukioyama0409@gmail.com]
\author{Masato Kamba}{NYX}[masato.kamba@nyx.foundation]
\author{Koshi Ota}{NYX}[cassiopeia01620@gmail.com]

\begin{abstract}
Since 2023, bug bounty platforms have introduced participation rules, such as
proof-of-concept mandates, against a flood of AI-generated
low-quality submissions. The platforms set these rules, but vendors running
programs on them bear the cost when a vulnerability goes unreported and gets
exploited. We ask whether the rules benefit the vendors they are meant to
protect. Adding to Gal-Or, Hydari, and Telang's 2024 model a vendor triage
cost per report reviewed and a participation cost paid in effort by the
researchers reporting vulnerabilities, we locate a threshold in an effective
triage-burden parameter, which captures the vendor's processing burden per
unit of aggregate search effort, with the rules lowering profit below it and
raising it above. Because attackers never use the platform, the
participation cost falls on researchers alone and fewer take part. Bug-finding
coverage falls and breach damage rises, which costs the vendor, while its
triage bill falls, which helps.
\end{abstract}

\begin{ekeyword}
bug bounty, vulnerability disclosure, security economics, platform policy
\end{ekeyword}

\maketitle

\begingroup
\renewcommand{\thefootnote}{}%
\footnotetext{This paper was presented at the Computer Security Symposium 2026
(CSS 2026) and published in its proceedings by the Information Processing
Society of Japan (IPSJ), October 2026. The copyright of this paper is held by
IPSJ. This version is posted with the permission of IPSJ, the copyright
holder.}%
\endgroup

\section{Introduction}\label{sec:intro}

Bug bounty programs, market mechanisms in which software vendors pay external
researchers for discovering and reporting vulnerabilities, have become a
standard component of vulnerability management at major firms~\cite{galor2024,sridhar2021}. Two developments have recently reshaped the cost structure
of this market. The broad availability of large language models and
AI-assisted code analysis tools has lowered the effort cost of vulnerability
discovery. Fang et al.~\cite{fang2024} show that LLM agents can autonomously
exploit one-day vulnerabilities, and
the volume of low-quality, often AI-generated submissions has risen with the
same tools~\cite{bugcrowd2026,stenberg2025}. Between 2023 and 2026, major bug bounty
platforms responded with participation rules: proof-of-concept mandates,
submission caps, deposits, and identity gates. The platforms set these rules,
but the vendors running programs on them bear the cost when a vulnerability
goes unreported and is exploited. We ask whether the rules benefit the
vendors they are meant to protect. The answer matters to any vendor deciding
how much to rely on a platform that imposes them.

We answer inside the game-theoretic framework of Gal-Or, Hydari, and Telang~\cite{galor2024}, which models the interaction among the vendor, white-hat
researchers, and black-hat attackers as a two-stage Stackelberg game and
derives the vendor's expected profit from operating a program. That framework
treats the platform as implicit. It represents neither the participation
costs the platform's rules impose on researchers nor any cost to the vendor
of processing the reports it receives. The same gap runs through the adjacent
theoretical literature~\cite{li2018,zhou2023}, while the empirical literature
characterizes platform dynamics descriptively without formalizing these costs~\cite{sridhar2021,zrahia2024}.

We add two components. The first is a participation cost that anti-spam
rules impose on white-hat researchers, together with a free-entry margin
under which a researcher joins only when the expected payoff from searching
covers this cost. The number of participating experts is then settled by the
researchers' own participation decisions. The second is a triage
cost on the vendor, the burden of processing the submissions the
program generates, which enters the vendor's profit alongside the damage and
bounty terms.

The central finding is that the friction moves vendor profit through two
channels. It falls on the white-hat researchers who protect the vendor by
finding and reporting severe bugs before attackers can exploit them, but it
does not reach the attackers, who never participate in the platform. As the
pool thins, attackers face weaker competition and succeed more often, so
coverage falls and expected breach damage rises. This channel always runs
against the vendor. The thinner pool also submits fewer reports, so the
triage bill falls. This channel always runs in its favor. Which channel
dominates depends on the vendor's effective triage burden, and the model
locates the
threshold at which the net effect changes sign. The threshold rises with the
vendor's breach cost, so anti-spam friction is most damaging to the vendors
with the most to lose. The result holds for the vendor whose posted bounty
stays fixed when the rules change. A vendor that
continuously re-optimizes its bounty never gains from the friction at
the calibrated parameter values, because
it can thin the pool itself, and its bounty response recovers
only part of the lost coverage.

\section{Background and Related Work}\label{sec:background}

\subsection{The Platform Layer and the AI Shift}\label{subsec:platform}

Most vendors do not run their bug bounty programs directly. They contract
with specialized platforms, including HackerOne, Bugcrowd, Immunefi,
Code4rena, Sherlock, Cantina, Intigriti, and YesWeHack.\footnote{HackerOne
(\url{hackerone.com}), Bugcrowd (\url{bugcrowd.com}), Immunefi
(\url{immunefi.com}), Code4rena (\url{code4rena.com}), Sherlock
(\url{sherlock.xyz}), Cantina (\url{cantina.xyz}), Intigriti
(\url{intigriti.com}), and YesWeHack (\url{yeswehack.com}).} These platforms
operate as two-sided markets, meaning they connect two groups that each
benefit from the other's presence, vendors on the demand side and researchers
on the supply side, and they provide submission intake, triage (the review
and validation of incoming reports), dispute resolution, and payment
processing~\cite{zrahia2024,wachs2022}. The institutional feature that
matters here is the division of control. The vendor specifies the program's
scope and bounty amounts, but the platform sets the participation rules that
determine which researchers can enter and which submissions reach the vendor.

The cost structure of vulnerability discovery shifted with the broad
availability of large language models. CVE publication reached record levels
in 2025~\cite{gamblin2026}, and by 2025 platforms reported that the same
tools flooded
their queues with low-quality submissions. GitHub stated that new tools,
including AI, have lowered the barrier to entry~\cite{githubsl2026}, and
Bugcrowd cited a 334-percent increase in queue volume over three weeks~\cite{bugcrowd2026}. The response is documented in a series of public policy
changes between 2023 and 2026. GitHub introduced stricter proof-of-concept
and scope-enforcement rules for its program in May 2026, requiring a working
demonstration of each reported bug and stricter adherence to the declared
program scope~\cite{githubsl2026}.
Immunefi capped newly registered researchers at one report per twenty-four
hours until their first paid report~\cite{immunefi2025}. Cantina added
per-bounty deposits, money a researcher stakes on each submission, which the
platform keeps for spam and low-effort AI reports and refunds on good-faith
rejections~\cite{cantina2026}. Code4rena introduced
signal-based submission caps, where a researcher's signal is a score the
platform computes from their track record of valid reports and a low score
means stricter limits~\cite{code4rena2025}. Sherlock has withheld
payouts since July 2023 from researchers below cumulative validity
thresholds, meaning a minimum count and share of valid reports across an
account's lifetime. This rule predates the AI flood~\cite{sherlock2023}. Bugcrowd announced
permanent bans in March 2026 for submission farming, the practice of
mass-submitting low-effort reports in the hope that some pay out~\cite{bugcrowd2026}.
HackerOne's Node.js program went furthest, blocking new researchers with no
signal from submitting at all~\cite{nodejs2026}. Two of these cases, GitHub's
and Node.js's, are program-level actions by a vendor, not
platform-level rules, and they show the same family of rules appearing on
both sides of the division of control.

What matters for the model is who pays these costs and who does not. Every
rule above raises the cost of taking part for researchers, and none of them
raises the cost for attackers, who never use the platform. Some rules bind
only new researchers, while deposit requirements, proof-of-concept mandates,
and payout-reserve thresholds bind established experts as well, and it is
this latter kind that the participation cost of Section \ref{sec:model}
stands for. The evidence that such costs reach capable researchers is direct.
GitGuardian reported leaked private keys closed as informative, a triage
label that acknowledges a report but pays no bounty, because triagers
demanded exploitation proofs that would have crossed into active
attack~\cite{gitguardian2026}, and the researcher-side literature documents
these burdens systematically~\cite{akgul2023}. The curl project ended its
bounty in January 2026 after an AI-generated report flood~\cite{stenberg2026}, and its maintainer had acknowledged months earlier that
even reducing the reward removes the traction for some professional and
highly skilled researchers~\cite{stenberg2025}.

\subsection{The Gal-Or, Hydari, and Telang Model}\label{subsec:galor}

Our analysis builds on the game-theoretic model of Gal-Or, Hydari, and Telang~\cite{galor2024}, restated here in the order the game is
played. There are four types of agents. The vendor releases the software and
pays the bounties. Expert white-hat hackers, $n$ of them, are researchers who
report what they find to the vendor in exchange for the bounty, and they can
search for both severe and non-severe bugs, where severe bugs are the ones
that map to the High or Critical ratings of the Common Vulnerability Scoring
System (CVSS). Non-expert white-hats, $l$ of
them, search only for non-severe bugs, since severe bugs are assumed too
costly for them per unit of effort. Black-hat hackers, $m$ of them, exploit
what they find for illicit gain, and they search only for severe bugs,
because non-severe bugs are assumed to be worth almost nothing to an
attacker.

The game is a Stackelberg game with two stages, meaning one side commits
first and the others respond. In the first stage the vendor commits to a
release
time $t$ and to two bounties, $p_s$ for a severe bug and $p_{ns}$ for a
non-severe one. In the second stage the hackers choose how hard to search,
and their choices form a Nash equilibrium given the vendor's commitment. The
race for each bug is what connects the two sides. Only the first valid report
of a bug is paid, and once a black-hat exploits a bug the vendor identifies
and patches it, so a bug that an attacker reaches first is closed to the
researchers. Searching is costly, with effort costs quadratic in effort and
severe bugs costing more per unit of effort than non-severe ones, which the
cost weights $c_w > 1$ for white-hats and $c_b > 1$ for black-hats capture.

How likely each side is to reach a bug first depends on everyone's effort.
Write $s \equiv n + m$ for the number of severe-bug searchers, $\alpha_s$ for
each expert's effort on severe bugs, and $\mu_s$ for each black-hat's effort,
where hackers of the same type choose the same effort in equilibrium. Each
hacker chooses its own effort taking all other hackers' efforts as given, and
the success probabilities below are evaluated at that type-symmetric
profile. The
probability that a given expert or a given black-hat is the first to a severe
bug is its own effort measured against the effort of its rivals in the race,
\[
\mathbb{P}_{ie}^{s} = \frac{(s-1)\,\alpha_s}{s\left[(n-1)\alpha_s + m\mu_s\right]},
\qquad
\mathbb{P}_{ib}^{s} = \frac{(s-1)\,\mu_s}{s\left[n\alpha_s + (m-1)\mu_s\right]} .
\]
This is the multiplicative ratio form of the model, given in their Appendix
A, and we work with it throughout because in the additive main-text form a
hacker's optimal effort does not respond to rivals' effort, so that form
cannot generate the cross-side substitution our mechanism runs on. We also
work within their severe-focus case, in which experts allocate effort only to
severe bugs.

Each hacker weighs the expected reward against the effort cost. An expert
chooses $\alpha_s$ to maximize
$K_s(t)\,\mathbb{P}_{ie}^{s}\,(r_s + p_s) - \tfrac{1}{2}c_w\alpha_s^2$, where
$r_s$ is the reputational gain from a valid report and $K_s(t)$ is the
probability that a severe vulnerability is still present when the software is
released at time $t$, decreasing and convex in $t$. A black-hat chooses
$\mu_s$ to maximize
$K_s(t)\,\mathbb{P}_{ib}^{s}\,W - \tfrac{1}{2}c_b\mu_s^2$, where $W$ is the
illicit gain from exploiting the bug, defined net of the legal risk of being
caught.

The vendor's expected profit is their Equation 14, which we write in full,
\begin{multline*}
\Pi = R(t) - K_s(t)\, m\, \mathbb{P}_{ib}^{s}\, TC_s
- K_s(t)\, n\, \mathbb{P}_{ie}^{s}\, p_s \\
- K_{ns}(t)\, l\, \mathbb{P}_{ine}^{ns}\, p_{ns}
- K_{ns}(t)\, TC_{ns} \left[ 1 - l\, \mathbb{P}_{ine}^{ns} \right] .
\end{multline*}
The first term $R(t)$ is the revenue from releasing at time $t$. The four
remaining terms are the costs, in order. The expected damage from black-hats,
since $m\,\mathbb{P}_{ib}^{s}$ of the remaining severe bugs are reached by
attackers first and each costs the vendor $TC_s$. The severe bounty bill,
since $n\,\mathbb{P}_{ie}^{s}$ of them are reported first by experts and each
earns $p_s$. The non-severe bounty bill, where $K_{ns}(t)$ is the probability
that a non-severe bug remains at release and $\mathbb{P}_{ine}^{ns}$ is the
probability that a given non-expert is the first to it. And the cost of
non-severe bugs that escape the program entirely and are discovered by users
instead, at $TC_{ns}$ each, with $TC_s$ much larger than $TC_{ns}$.

Two of their results anchor ours. Raising the severe bounty $p_s$ raises
white-hat success and lowers black-hat success, because the two sides race
for the same bugs (their Proposition 1). And a program with positive optimal
bounties is more profitable than no program (their Proposition 2).

\subsection{The Gap}\label{subsec:gap}

B\"{o}hme~\cite{boehme2005} establishes vulnerability information as a
tradeable economic object in a market with structural failures, and Kannan
and Telang~\cite{kannan2005} show that an unregulated vulnerability market
can perform worse than coordinated disclosure. On the empirical side, Sridhar
and Ng~\cite{sridhar2021} estimate from over 3{,}800 HackerOne program-months
a low median supply elasticity with respect to bounty
amounts, interpreting researchers as motivated mainly by things other than
money.
Zrahia, Gandal, Markovich, and Riordan~\cite{zrahia2024} treat the platform
as a two-sided marketplace and show that the COVID-19 supply expansion
produced a disproportionate increase in duplicate valid submissions, diluting
the expected reward for valid work. Allodi~\cite{allodi2017} documents that
underground markets price exploits at levels comparable to or above
legitimate bounties and refresh their stock slowly. Closer to the platform
layer, Wachs~\cite{wachs2022} analyzes platforms as transaction-cost
intermediaries that make the market possible, and Zhao, Laszka, and Grossklags~\cite{zhao2017} measure platform ecosystems and evaluate policy instruments
for them. None of these formalizes how participation costs imposed by the
platform itself affect the market equilibrium. Laszka, Zhao, and Grossklags~\cite{laszka2016}
model invalid-report validation and organizational processing costs in
bug-bounty platforms, so they come closest to pricing report handling. Their
model has no race for severe bugs between researchers and attackers. Ours
differs by embedding a vendor-side triage burden and a free-entry
participation margin into that race, which is what lets us study the opposing
effects of participation friction on aggregate coverage and triage
expenditure. The two theoretical contributions closest to
ours share the same
gap. Li and Liao~\cite{li2018} model defender--attacker interaction without a
platform layer, and Zhou, Ma, and Feng~\cite{zhou2023} model a platform's
decision whether to launch a program for third-party software, in which the
participation costs a running program imposes on researchers are not the
object. Within the race model we build on, the vendor's profit function
counts only successful discoveries and carries no processing cost.
This paper addresses both omissions.

\section{Model}\label{sec:model}

\subsection{Friction as a Single Lever}\label{subsec:lever}

We abstract the family of anti-spam rules into a single parameter, a
participation cost $c_f \geq 0$ borne by white-hat researchers, which we call
the friction. The cost need
not be monetary. It may be the time required to produce a working proof of
concept, a deposit placed at risk, the disclosure of one's identity, or a cap
on how many reports one may submit. What these policies share is that they
raise the burden of taking part, and $c_f$ is the common measure of that
burden. The bundled burdens differ in kind and in who bears them, since a
deposit binds harder on a researcher without conventional banking and an
identity gate excludes researchers in some jurisdictions outright~\cite{akgul2023}. The parameter acts through a single margin, the size of
the expert white-hat pool, and we treat the platform's choice of $c_f$ as an
input. The AI-era surge in
submission volume that motivated the rules is context for the analysis, not a
variable in it.

\subsection{The Participation Margin}\label{subsec:entry}

This section derives how the researcher pool is determined. The order of
moves has three steps. The vendor posts its bounty, researchers observe the
bounty and the friction and enter until the marginal researcher is
indifferent, and the hackers then choose effort taking the pool as settled.
Write $\pi_W(n; p_s)$ for an expert's expected payoff from taking part
when the pool holds $n$ experts at the posted bounty $p_s$, the expected
reward for a valid report on a remaining severe bug, minus the effort cost of
searching (the exact
expression is in
Appendix \ref{app:derivations}). Following the standard free-entry
formulation of~\cite{mankiwwhinston1986}, a researcher joins when the payoff
covers the friction, so the marginal researcher is indifferent,
\[
\pi_W(n^{\ast}; p_s) \;=\; c_f .
\]

This condition picks out a single pool size at each posted bounty, and
raising the friction shrinks
it. Adding an expert splits the same prize among more rivals, so the payoff
$\pi_W$ falls as the pool grows. We show this
monotonicity analytically up to a stated, negligible exception (Appendix
\ref{app:entry}). When the platform raises $c_f$, the payoff at the incumbent
pool no longer covers the friction, so researchers leave until the margin is
restored at a smaller $n^{\ast}$. The pool empties only when the friction
reaches the payoff of a lone searcher, $c_f \geq \pi_W(1; p_s)$.
One behavioral assumption matters for what follows, and we state it as an
assumption, not a fact. The posted bounty is set for coverage, without
the triage burden priced in, and it stays posted when the platform changes
its rules. Section \ref{subsec:active} examines the
vendor that continuously re-optimizes its bounty instead, and the answer
changes there.

The friction reaches only one side of the market. Attackers never use the
platform, because exploitation requires no submission, so they bear none of
the friction and $m$ stays fixed. Two qualifications bound this premise.
Identity-based instruments reach anyone maintaining a platform identity, so
an attacker who also holds a researcher account pays part of the friction.
The asymmetry is exact for effort, deposit, and workload costs, and
approximate for identity gates. The premise also needs attackers to find the
legitimate channel unattractive at the margin, and the calibration supports
this. Underground
pricing comparable to legitimate bounties with a slow refresh cycle~\cite{allodi2017} means a marginal attacker gains little by switching sides
for a bounty the vendor keeps small. Everything below depends on this
asymmetry.

One exact consequence of the entry margin is worth stating in words, because
Section \ref{subsec:sticky} uses it. At equilibrium an expert's gross reward
equals exactly twice the effort cost of searching, so the payoff from taking
part equals the effort cost itself, and a
higher friction makes each surviving researcher search \emph{harder}, even as
the pool thins. The derivation is in Appendix \ref{app:entry}. Aggregate
coverage $n\alpha_s$ still falls, and it is the aggregate that protects the
vendor.

With the pool size settled by the researchers' own participation decisions,
the vendor's problem gains the second new ingredient, the cost of processing
the reports the pool submits.

\subsection{The Vendor's Problem with Triage}\label{subsec:triage}

Reports have no representation in Gal-Or, Hydari, and Telang's game, which
counts only successful discoveries, so nothing in the equilibrium corresponds
to the flow of reports a vendor must
process. Pricing triage therefore
requires an assumption. We assume that the report flow scales with the pool's
aggregate search effort $n\alpha_s$. The coefficient $\beta$ captures the
effective triage burden per unit of aggregate search effort, combining the
processing cost of a report with the number of reports one unit of search
effort generates. We keep $\beta$ at this level of abstraction because the
two components cannot be told apart in the model. This is an
assumption, not a derivation, and Section \ref{sec:discussion} examines it.
Two of its properties can be stated without new computation. First, because
$\beta$ aggregates the processing burden and report-generation intensity into
a single model parameter, the location of the threshold of Section
\ref{subsec:threshold} depends on numbers only the model defines. Its existence and its
comparative statics are what the results say about the real world.
Second, the linear
form is conservative for the coverage result in a specific sense. If the
true flow is convex in the pool, as the duplicate-dilution evidence of~\cite{zrahia2024} suggests, the triage saving from a pool cut is larger than
the linear estimate, so the reported threshold is an upper bound, while the
region where coverage dominates at low $\beta$ is unaffected, since the
triage term vanishes with $\beta$ under any curvature.
The vendor's expected cost has three terms,
\begin{gather*}
C \;\equiv\; K_s(t)\left[\,m\,\mathbb{P}_{ib}^{s}\,TC_s \;+\; n\,\mathbb{P}_{ie}^{s}\,p_s\,\right] \;+\; \beta\,n\,\alpha_s , \\
\Pi = R - C - \text{const},
\end{gather*}
where const collects the non-severe terms of the profit function, which do
not depend on $n$. The vendor's other instruments, the release time $t$ and
the non-severe bounty $p_{ns}$, are held at their optimal values throughout.
The three terms of $C$ are the three ways the pool costs
the vendor money. The first is the expected black-hat damage. The second is
the severe bounty bill. The third is the new triage term, the cost of
processing the reports the pool submits.
We evaluate the friction's effect under two vendor behaviors, both computed
in Appendix \ref{app:derivations}. The posted-bounty vendor sets its bounty
once, as the optimum when the triage burden is not priced, and leaves it
posted when the friction changes. The re-optimizing vendor re-solves its
bounty after the rule change, anticipating how the bounty steers entry
through the participation condition of Section \ref{subsec:entry}. The main
analysis uses the first. Section \ref{subsec:active} uses the second.

\section{Results}\label{sec:results}

\begin{figure}[t]
\centering
\begin{tikzpicture}
\begin{axis}[
  width=\columnwidth, height=3.8cm,
  xlabel={platform participation friction $c_f$},
  ylabel={equilibrium pool $n^{\ast}$},
  xmin=0.10, xmax=0.20, ymin=0, ymax=45,
  label style={font=\small}, tick label style={font=\footnotesize},
]
\addplot[black, mark=*, mark size=1.2pt] coordinates
  {(0.100,43.23) (0.106,39.14) (0.111,35.45) (0.117,32.09) (0.122,29.02)
   (0.1285,25.81) (0.133,23.61) (0.139,21.21) (0.144,18.97) (0.150,16.88)
   (0.155,14.93) (0.161,13.08) (0.166,11.34) (0.172,9.69) (0.1750,8.79)
   (0.183,6.59) (0.188,5.11) (0.194,3.67) (0.200,2.23)};
\end{axis}
\end{tikzpicture}
\caption{As platform participation friction $c_f$ rises, the equilibrium
expert pool $n^{\ast}$ falls along the entry margin of Section
\ref{subsec:entry} at the posted bounty $p_s = 20.46$ of the base
calibration.}
\label{fig:participation}
\end{figure}
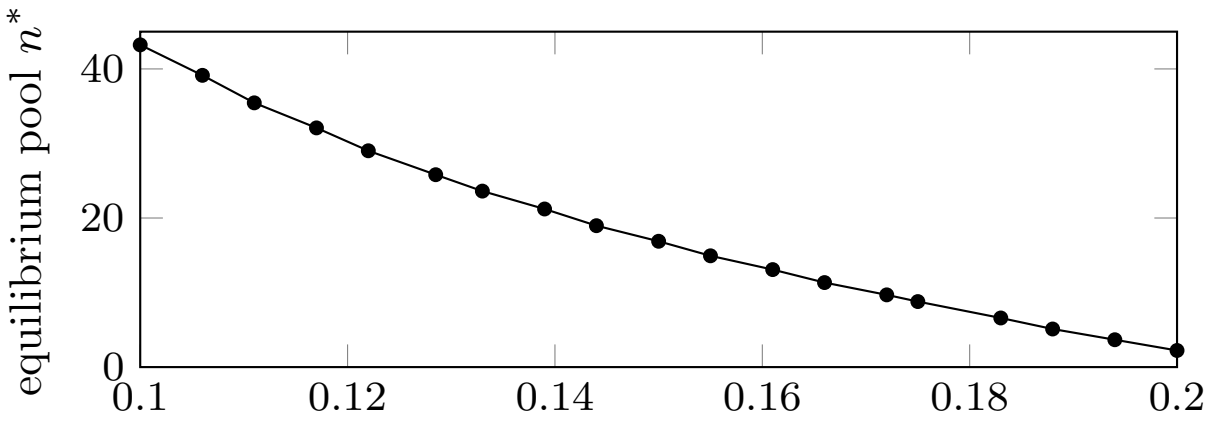

The results follow the friction in two steps, first into the size of the
researcher pool and then into the vendor's profit, and Figure
\ref{fig:participation} shows the first step.

\subsection{A Threshold in the Triage Cost}\label{subsec:threshold}\label{subsec:sticky}

\begin{figure}[t]
\centering
\begin{tikzpicture}
\begin{axis}[
  width=\columnwidth, height=3.6cm,
  xlabel={effective triage burden $\beta$},
  ylabel={$\Pi(c_f^{lo}) - \Pi(c_f^{hi})$},
  xmin=-0.03, xmax=0.88, ymin=-3.5, ymax=1.5,
  xtick={0,0.2,0.4,0.6,0.8},
  label style={font=\small}, tick label style={font=\footnotesize},
]
\addplot[black, mark=*, mark size=1.5pt]
  coordinates {(0,1.177) (0.1,0.620) (0.2,0.062) (0.3,-0.495) (0.4,-1.053)
  (0.5,-1.611) (0.6,-2.168) (0.7,-2.726) (0.8,-3.283)};
\addplot[dotted, thick, black]
  coordinates {(-0.03,0) (0.88,0)};
\addplot[dashed, black]
  coordinates {(0.211,-3.5) (0.211,1.5)};
\node[font=\footnotesize, anchor=west] at (axis cs:0.02,1.25) {friction hurts};
\node[font=\footnotesize, anchor=east] at (axis cs:0.86,-2.1) {friction pays};
\node[font=\footnotesize, anchor=west] at (axis cs:0.216,0.72) {$\beta^{\ast} \approx 0.211$};
\end{axis}
\end{tikzpicture}
\caption{The profit advantage of the low-friction state over the
high-friction state, at nine values of the effective triage burden $\beta$,
base calibration, posted-bounty vendor. The points at the five tabulated
burdens are exact differences of
the Table \ref{tab:sticky} values, and the remaining four lie on the same
line. Above zero the friction hurts the vendor.
Below zero it pays. The dashed line marks the threshold
$\beta^{\ast} \approx 0.211$.}
\label{fig:threshold}
\end{figure}
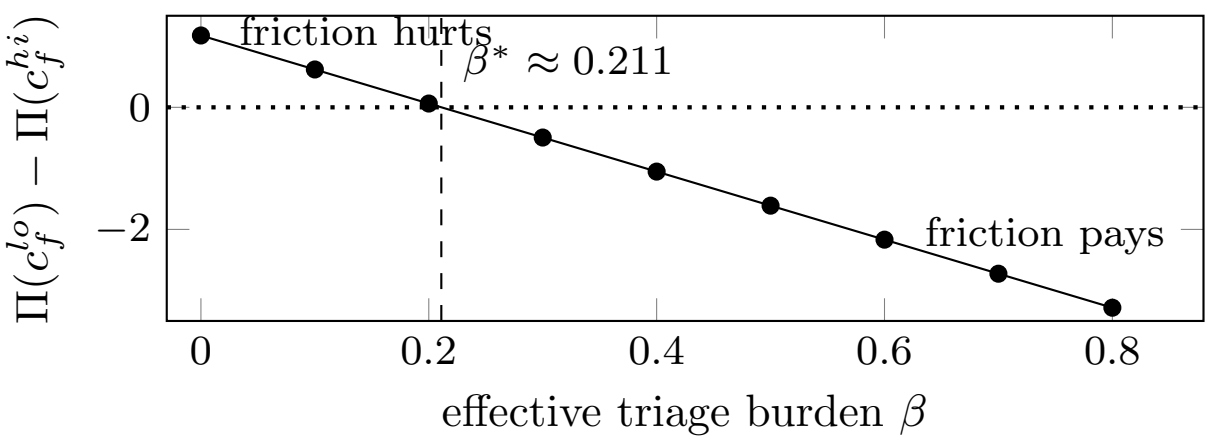

Whether the rules help or hurt the vendor depends on the effective triage
burden $\beta$. The units caveat of Section
\ref{subsec:triage} applies here. Table \ref{tab:sticky} evaluates the
equilibrium at the base calibration
($m = 10$, $K_s(t) = 0.5$, $c_w = c_b = 2$, $r_s = 10$, $W = 96$,
$TC_s = 30$, an example parameterization chosen for comparability with
the source model's numerical examples, not from micro-data), under a rule
change that raises the friction from $c_f = 0.1285$ to $c_f = 0.1750$,
values chosen so the pool falls from $25.8$ to $8.8$ at the posted bounty
$20.46$.
At zero triage burden the friction hurts. The thinner pool raises the
expected damage, and there is no triage bill to save. The vendor's cost
rises by $1.18$. As $\beta$ rises, the triage saving from the thinner pool
grows linearly, while the added damage does not depend on $\beta$. The two
cross at $\beta^{\ast} \approx 0.211$, below which the friction costs the
vendor and above which it pays. This is the break-even
point of the calibrated numerical model, expressed in effective
triage-burden units, not a directly measured dollar cost per report. Figure \ref{fig:threshold} shows the profit
difference.

\begin{table}[t]
\centering
\caption{Equilibrium under the friction rise, posted-bounty vendor, base
calibration. The pool falls from $25.81$ to $8.79$ at the posted bounty
$20.46$ in every row. $\Delta C$ is the cost change from the friction rise,
so a positive value means the friction hurts.}
\label{tab:sticky}
\footnotesize
\setlength{\tabcolsep}{3.2pt}
\begin{tabular}{lrrrrr}
\toprule
$\beta$ & $C(c_f^{lo})$ & $C(c_f^{hi})$ & $\Delta C$ & $\Pi(c_f^{lo})$ & $\Pi(c_f^{hi})$ \\
\midrule
$0.00$ & $13.100$ & $14.277$ & $+1.177$ & $36.900$ & $35.723$ \\
$0.20$ & $14.950$ & $15.013$ & $+0.062$ & $35.050$ & $34.988$ \\
$0.40$ & $16.801$ & $15.748$ & $-1.053$ & $33.199$ & $34.252$ \\
$0.60$ & $18.651$ & $16.483$ & $-2.168$ & $31.349$ & $33.517$ \\
$0.80$ & $20.502$ & $17.218$ & $-3.283$ & $29.499$ & $32.782$ \\
\bottomrule
\end{tabular}
\end{table}

The threshold is not a feature of one parameterization. Across thirteen
perturbations of the base calibration, spanning $W \in \{48, 200\}$,
$r_s \in \{2, 30\}$, $TC_s \in \{10, 100, 200\}$, $K_s \in \{0.2, 0.9\}$, both
cost multipliers, and a joint shift in $W$ and $r_s$, a positive threshold
exists in all fourteen parameter sets, so the coverage channel dominates at
low triage burden throughout (Appendix \ref{app:robustness}). Two mechanisms
produce the flips, and they should not be pooled. In nine of the fourteen
the thinned pool survives at the posted bounty, and the triage saving, linear
in $\beta$, overtakes the fixed damage penalty. In five
($W = 48$, $r_s = 30$, $TC_s = 100$, $TC_s = 200$, $c_w = 1.2$) the posted
bounty cannot sustain the thinned pool under the higher friction, the pool
collapses, and the flip compares a low-friction state whose triage bill has
grown past the cost of having no program at all, the collapsed
high-friction state with unmitigated damage at $n^{\ast} = 0$. Our claim
rests on the first mechanism, and the second we report but do not emphasize.

\subsection{When the Vendor Re-Optimizes}\label{subsec:active}

The threshold above belongs to the posted-bounty vendor. The other behavior
answers differently. A vendor that re-solves its bounty after the rule
change, anticipating how the bounty steers entry, has no region where the
friction pays at the base calibration. The cost increase from the same friction rise falls
from $+0.92$ at $\beta = 0$ to exactly zero at
$\beta_0 = 0.3024$ and
never turns negative, on a $0.001$ grid refined by bisection, because any
thinning the vendor wants it can produce by lowering its own bounty, so the
friction, which makes every pool size
costlier to sustain, has nothing to offer. Past $\beta_0$ the optimal bounty
attracts no entry and the program effectively shuts down. We do not read realism into that
case, since real programs degrade gradually and vendors hold margins our
model freezes. One exception survives this shutdown result: when the friction is
low enough that even a zero bounty attracts entry, the vendor cannot thin the
pool itself, and a genuine flip survives (at $TC_s = 10$ with
$c_f^{lo} = 0.017$, $\beta^{\ast} = 0.2287$).

The re-optimizing vendor's bounty response recovers part of the coverage but
not the profit. Table \ref{tab:active} reports the equilibrium at
$\beta = 0.20$. The optimal bounty rises from $17.34$ to $19.67$, and
preserves $42$ percent of the pre-change pool, which would otherwise collapse
to zero at the old bounty. The cost still rises, from $14.558$ to $14.993$.
The bounty buys back some of the pool. It does not buy back the loss.

\begin{table}[t]
\centering
\caption{Equilibrium under the friction rise, re-optimizing vendor, base
calibration at $\beta = 0.20$. Entries use the unrounded
friction values ($c_f^{lo} = 0.12846$, $c_f^{hi} = 0.17500$).}
\label{tab:active}
\footnotesize
\setlength{\tabcolsep}{3.2pt}
\begin{tabular}{lrrrrrrr}
\toprule
 & $p_s^{\ast}$ & $n^{\ast}$ & $\alpha_s$ & $\mu_s$ & $\mathbb{P}_{ie}^{s}$ & $C$ & $\Pi$ \\
\midrule
$c_f^{lo}$ & $17.338$ & $14.989$ & $0.3584$ & $1.3291$ & $0.0188$ & $14.558$ & $35.442$ \\
$c_f^{hi}$ & $19.673$ & $6.318$ & $0.4183$ & $1.4421$ & $0.0236$ & $14.993$ & $35.007$ \\
\bottomrule
\end{tabular}
\end{table}

There is a reason a platform might not see the harm. As the pool thins, the
surviving experts face fewer rivals and, by the identity of Section
\ref{subsec:entry}, each searches harder, so each wins races more often.
The per-hunter success probability $\mathbb{P}_{ie}^{s}$
rises from $0.0169$ to $0.0230$ at the posted bounty. Per-hunter dashboards, mean reward per active
researcher, and win rates all improve even as aggregate coverage falls. A
policy that harms the hosted vendor can look, from the platform's
vantage, like one that is working.

\subsection{High-Stakes Vendors Lose Most}\label{subsec:stakes}

For the posted-bounty vendor, the stakes decide the size of the loss and
the threshold's location. At
zero triage burden the cost increase from the friction rise is $+0.51$ at
$TC_s = 10$, $+1.18$ at the base $TC_s = 30$, and $+11.01$ at $TC_s = 100$,
and the threshold rises with the stake, from $\beta^{\ast} = 0.118$ at
$TC_s = 10$ to $0.211$ at the base and $0.849$ at $TC_s = 100$, reaching
$1.92$ at $TC_s = 200$ (Appendix \ref{app:robustness}). No clean
proportionality between $\beta^{\ast}$ and $TC_s$ emerges here.
At the two highest stakes in the sweep, $TC_s = 100$ and $TC_s = 200$, the
flip is of
the collapse type, so the benefit is measured against a low-friction
state whose triage bill has outgrown the cost of having no program at all. The
high-stakes thresholds are real but read differently from the
thinning flips below them.
Anti-spam friction is therefore most damaging to the vendors with the most to
lose, and they stay on the losing side over
the widest range of triage burdens.

\subsection{Interpreting the Result}\label{subsec:meaning}

The result is counter-intuitive in one respect. Below the threshold the
vendor spends less under the friction, yet is worse off. The accounting is
direct under a posted bounty. The friction rise pays out fewer bounties,
since fewer researchers win, and it cuts the triage bill, since fewer reports
arrive. What grows is the
expected damage, because the attackers face weaker competition. At low
$\beta$ the damage increase dominates and the vendor loses. The saving grows
with $\beta$ and the damage does not, which is why the sign flips exactly
once.

The vendor's private loss understates the social cost. The bounty the vendor
saves is a transfer whose disappearance is socially neutral. The added
black-hat damage is a real loss borne by the vendor and, through the breach
risk it creates, by the software's users. The private profit decline is therefore
a lower bound on the social cost of the friction at the coverage margin. What
anti-spam delivers in return is triage efficiency, which adds no bug-finding
coverage on its own. A filter
that removes only the reports that were never going to protect the vendor,
and deters no genuine researcher, captures
the same triage saving while giving up no coverage, so it dominates a
participation barrier that buys the saving by thinning the pool. That is a
strong requirement, a condition, not a
description of what platforms have available. A wrongful rejection is itself
a stochastic participation cost through the entry margin of Section
\ref{subsec:entry}, so the dominance holds exactly for filters whose
false-positive rate on genuine reports keeps the added expected friction
below the saving they capture. Blunt barriers persist because
building a filter that precise is hard, as Section \ref{subsec:platform}'s
GitGuardian episode shows.

\subsection{What a Vendor Can Evaluate}\label{subsec:operational}

The vendor's cost change decomposes exactly as
$\Delta C = TC_s\,\Delta B - \text{savings}$, where
$B = K_s(t)\,m\,\mathbb{P}_{ib}^{s}$ is the mass of severe bugs won by
black-hats and the savings term, the fall in triage and bounty outlay, is in
the vendor's own records. Those records show report volumes,
processing times, and bounty outlays, and the effective triage burden $\beta$
itself cannot be estimated separately without an estimate of how many reports
one unit of search effort generates. Define the substitution rate $\sigma$ as the ratio
of the added black-hat mass to the drop in valid severe reports. No more than
the lost reports can be picked up by attackers, so as a statement about the
world, $\sigma \leq 1$.\footnote{The ratio-form contest
inherited from~\cite{galor2024} does not impose the bound internally, since
its unnormalized first-finder shares sum to slightly more than one.
What makes the test convincing is the definition, not the model
specification.} The added damage is then at most $TC_s$ times the drop, so
the condition
\[
\text{savings} \;>\; TC_s \times (\text{drop in valid severe reports})
\]
is sufficient for the rule to have helped. It is one-sided, since it stays
silent in many cases where the rule did help, but it never says the rule
helped when it did not. The coverage loss is a movement
in the vendor's own report volume, detectable within tens of
observed events ($13$ for the report-volume drop in the friction rise of
Table \ref{tab:sticky}, a rate ratio of $2.52$). The damage change is a movement in breach
counts, and detecting it there requires an order of
magnitude more ($112$ at the base calibration, a rate ratio of $1.41$), which few vendors
accumulate quickly. Both figures come
from a two-sided Poisson rate-ratio power calculation at $\alpha = 0.05$ and
power $0.80$ over equal windows, against baselines of $9.25$ reports and
$0.058$ breaches per window, and are lower bounds,
since confounding, attribution error, and detection lag worsen the
requirement. A vendor's records therefore settle
the savings side and the coverage loss. The damage side requires an
estimate of what an exploit is worth to an attacker, which no record
contains.

\section{Discussion and Limitations}\label{sec:discussion}

\subsection{Relation to the Empirical Literature}\label{subsec:empirical}

The friction analysis connects to the elasticity estimate of Sridhar and Ng~\cite{sridhar2021} in two ways. A low median supply elasticity indicates
that the reputational component $r_s$, not the bounty, is the
part of researcher compensation that matters, so a policy cost that erodes the
participation payoff attacks the margin the empirical literature identifies
as dominant. The same estimate is the empirical counterpart of Section
\ref{subsec:active}, since even large bounty increases produce little
additional supply, so vendors cannot be expected to undo friction through
prices. Zrahia et al.~\cite{zrahia2024} document that
researchers' net compensation moves with platform conditions, since the
COVID-19 supply expansion diluted expected rewards through duplicates, and
friction operates on the same object, the compensation a researcher expects
net of participation costs. Their two-sided framing explains
its institutional origin, since the platform controls
the participation rules, not the bounties. Allodi~\cite{allodi2017} supports the
asymmetry behind the result. Underground markets price exploits at
levels comparable to or above legitimate channels and refresh slowly,
consistent with treating $W$ as stable over the horizon on which platforms
adjust their rules.

\subsection{Limitations}\label{subsec:limitations}

\paragraph{The model has no submission count of its own.} Gal-Or, Hydari, and
Telang count only successful discoveries, so bringing triage into the
vendor's profit required two imposed assumptions. We assumed that reports
scale with
search effort, and that every report costs the same to process. Accordingly,
$\beta$ is an effective burden coefficient, not a directly observed cost per
report, and the model does not distinguish the processing cost of low-quality
and valid reports, nor can the report-generation rate be estimated
separately. The second assumption is
the more vulnerable, since dismissing an obviously worthless report is cheaper
than verifying a real finding. Section \ref{subsec:triage} bounds the first
assumption's curvature. The deeper omission is that the modeled flow contains
no low-quality reports. The rules of Section \ref{subsec:platform} were meant
to deter spam while keeping genuine researchers, and that benefit never
enters the model's ledger.

\paragraph{The effective triage burden is an input, and only the defender side is
recoverable.} Which side of the threshold a vendor occupies depends on its
own effective triage burden, which we treat as a parameter and do not estimate. The
identities of Section \ref{subsec:entry} cut both ways here. The pool size,
individual effort, and friction level on the defender side are recoverable
from the vendor's valid-report counts and the platform's published policy,
but the damage term depends on the attacker's effort, cost, and gain, none of
which any record contains. The direction of the loss is identified, but its
level is not.

\paragraph{The posted bounty stays posted.} Our main analysis holds the
bounty fixed when the platform raises the friction, set for coverage without
the triage burden priced in. A vendor that re-optimizes its bounty against
the entry margin instead faces no region where the friction pays at the
base calibration,
because it can thin the pool on its own (Section \ref{subsec:active}), and
the alternative order in which the vendor prices against a given pool
produces yet another threshold location (Appendix \ref{app:alttiming}). The
answer therefore depends on vendor behavior, and the reader's case is the
behavioral question of how quickly real bounty tables move. One
caveat cuts against the re-optimizing reading too, since its argument grants the
vendor control of report volume through its bounty, while the volume surge
followed the availability of AI tools, not any change in bounties~\cite{githubsl2026,bugcrowd2026}.

\paragraph{The analysis stays in the severe-focus case and in one
vendor--platform pairing.} Friction large enough to push experts toward mixed
effort allocations is outside scope, and researchers work on several
platforms at once, moving toward better-paying programs~\cite{maillart2017}, so the result concerns a given pairing, not the
market-level outcome. The friction also binds on the non-expert pool in
reality, and the non-severe terms would move with it in the same direction.
We hold those terms constant. By the same token, the persistence of friction regimes
across the platforms of Section \ref{subsec:platform} suggests the rules
deliver value the model
excludes, since vendors are the platform's paying side and can re-list
elsewhere.

\paragraph{The hurts-versus-helps answer can flip with the chosen numbers,
but the threshold always exists.} At the base
calibration the attacker gain $W = 96$, about three times the researcher's
total reward, puts black-hats first in most races, against measured
exploitation rates of a few percent of disclosed vulnerabilities~\cite{jacobs2020}. Recalibrating $W$ downward to plausible first-find shares
lowers the threshold with it, to between $0.03$ and $0.20$ across shares of
$5$ to $50$ percent, so the verdict at the chosen numbers moves from
hurts to helps. A positive threshold exists at every share tested, but the
sign at any particular $\beta$ changes with $W$. First-to-find is not
exploited-in-the-wild, so the implied $W$ and the recalibrated thresholds are
lower bounds. The search-space premise is the second weak point. Every result
inherits from Gal-Or, Hydari, and Telang the premise that both sides race for
the same severe bugs, yet RAND attributes its low collision rates in part to
the two sides looking for different vulnerabilities with different techniques~\cite{ablonbogart2017}, and Atefi et al.\ find that exploited and
hunter-reported vulnerabilities differ systematically~\cite{atefi2023}.
Measured collision rates climb with the observation window, from under one
percent at ninety days to about forty percent over fourteen years in RAND's
data, with public-codebase estimates in between~\cite{finifter2013,herr2017},
so the objection weakens as the window grows. If the two sides search largely
disjoint spaces, or if the race runs on a short horizon, the coverage-loss
channel is weaker than modeled. What both weak points put at risk is the hurts-versus-helps
verdict at the base calibration of Section \ref{subsec:threshold}, not the existence
of the threshold. Finally, the empirical estimates we rely on predate the AI
era, since the Sridhar and Ng and Zrahia et al.\ samples end in 2020 and
2021, so their use here assumes their direction has persisted.

\section{Conclusion}\label{sec:conclusion}

Platform anti-spam rules reach the vendor through two channels. They thin the
white-hat researchers who protect the vendor while leaving the attackers, who
never participate in the platform, untouched, so coverage falls and breach
damage rises. They also shrink the flow of reports the vendor pays to process,
so the triage bill falls. Which channel dominates depends on the vendor's
effective triage burden, which combines the processing burden of incoming
reports with the intensity at which search effort generates reports. Within
the tested parameter range the model
locates the calibrated threshold at which the sign of the
profit effect changes, for the vendor whose posted bounty does not re-adjust
when the rules change. For a vendor below the
threshold the outcome is a false economy, since it spends less on bounties and less
on triage yet is worse off, because the added damage exceeds both of those
savings. The harm is easy for a platform
to miss, since per-hunter metrics improve as the pool thins. And the vendor's
private loss, where it occurs, is a lower bound on the social cost. The
comparison favors filters that target low-value submissions directly over
blunt participation barriers, provided their false-positive rate
stays low enough that the added expected friction does not eat the saving,
and vendors should price a platform's friction into their reliance on it,
since even an optimized bounty response recovers only part of the coverage.
The natural next step is empirical. A vendor's report
volume and processing times are measurable from its records, but
locating itself
against the threshold requires an estimate of the attacker's
gain, which no record contains.

\section*{Acknowledgments}

We thank Professor Kazue Sako and Professor Tatsuya Mori of Waseda
University, and the members of
their laboratories, for their support of this research through
reviews and many discussions of this paper.

\bibliographystyle{unsrt}
\bibliography{reference_main}

\appendix
\section{Derivations and Robustness}\label{app:derivations}

This appendix restates the derivations the main text defers, working in the
multiplicative ratio form of Gal-Or, Hydari, and Telang's Appendix A~\cite{galor2024} and in the severe-focus case. The equilibrium is a coupled
nonlinear system with no closed form, so the derivations stop at the defining
first-order conditions and the entry fixed point, and the reported quantities
are obtained by solving those conditions numerically. The procedure is
deterministic, and all reported numbers are reproducible from the displayed
equations. An independent implementation from the first-order conditions
returns the same values (at the base calibration with $\beta = 0.20$,
$C = 14.9503$ at the low-friction value and the threshold
$\beta^{\ast} = 0.2111$). One property of the ratio form should be noted
here, since everything below builds on it: the first-finder shares do not sum
to one, since each type's denominator omits the focal hacker's own effort.
Their total mass runs slightly above one. This is inherited from the source
model and affects none of the sign conclusions.

\subsection{Stage 2: Equilibrium Effort}\label{app:stage2}

For the focal expert white-hat (eWHH) and the focal black-hat (BHH), evaluated at the type-symmetric
profile, the success probabilities are as in Section \ref{subsec:galor}.
Differentiating the objectives and imposing symmetry yields the coupled
equilibrium conditions
\begin{gather*}
c_w\,\alpha_s\left[(n-1)\alpha_s + m\mu_s\right] = \kappa\,(r_s + p_s), \\
c_b\,\mu_s\left[n\alpha_s + (m-1)\mu_s\right] = \kappa\,W ,
\end{gather*}
with $\kappa \equiv K_s(t)(s-1)/s$. The solver evaluates them by multi-start
root finding and retains the root continuous with the frictionless
equilibrium. Across every parameter set tested, the search returned no second
root with positive efforts. The eWHH equilibrium payoff used in Section
\ref{subsec:entry} is
\[
\pi_W(n; p_s) = K_s(t)\,\mathbb{P}_{ie}^{s}\,(r_s + p_s) - \tfrac{1}{2}\,c_w\,\alpha_s^2 ,
\]
and the identity
$\pi_W = \tfrac{1}{2}c_w\alpha_s^2$ follows by substituting the eWHH
condition into the payoff, and holds to machine precision in every solver
state.

\subsection{Stage 1: The Vendor's Optimal Bounty}\label{app:stage1}

The vendor commits to $p_s$ before entry, so its Stage 1 problem is
\[
\min_{p_s \geq 0}\; C\!\left(p_s,\, n(p_s; c_f)\right),
\]
where $n(p_s; c_f)$ is the entry response of Section \ref{subsec:entry}. The
first-order condition is
\[
\frac{\mathrm{d}C}{\mathrm{d}p_s}
\;+\; \frac{\partial C}{\partial n}\,\frac{\mathrm{d}n}{\mathrm{d}p_s} = 0 ,
\]
where the first term is the total derivative through the Stage 2 system,
\begin{multline*}
\frac{\mathrm{d}C}{\mathrm{d}p_s}
= K_s(t)\left[\,m\,TC_s\,\frac{\mathrm{d}\mathbb{P}_{ib}^{s}}{\mathrm{d}p_s}
+ n\,\mathbb{P}_{ie}^{s}
+ n\,p_s\,\frac{\mathrm{d}\mathbb{P}_{ie}^{s}}{\mathrm{d}p_s}\,\right] \\
+ \beta\,n\,\frac{\mathrm{d}\alpha_s}{\mathrm{d}p_s} ,
\end{multline*}
and the steering term follows from the entry condition as
$\mathrm{d}n/\mathrm{d}p_s = -(\partial \pi_W / \partial p_s)\,/\,(\partial \pi_W / \partial n)$,
with all derivatives obtained by implicitly differentiating the
Stage 2 conditions, a step Gal-Or, Hydari, and Telang's
Appendix A leaves undone. The posted bounty of the main text is the optimum
of this problem at $\beta = 0$ at the low-friction value, since the posted
bounty is set without the triage burden priced in. The re-optimizing vendor
of Section \ref{subsec:active} solves the full problem at each
$(\beta, c_f)$. The solver locates the optimum by a bounded
search guarded by a coarse grid against non-unimodality, with no second local
minimum found in practice, and the points where the non-negativity
constraint binds are reported with the results.

\subsection{The Entry Margin}\label{app:entry}

The property that gives the entry margin its structure is that
$\pi_W(n; p_s)$ is
strictly decreasing in $n$ at any posted bounty. The solver confirms
$\pi_W'(n) < 0$ over the operating
range at every bounty tested. More can be said analytically. By the identity of Appendix
\ref{app:stage2}, $\pi_W = \tfrac{1}{2}c_w\alpha_s^2$ exactly, so the sign
of $\pi_W'(n)$ at a fixed bounty is the sign of $\alpha_s'(n)$. Total
differentiation of the coupled Stage 2 conditions gives
$\mathrm{d}\alpha_s/\mathrm{d}n = c_w c_b\,\alpha_s N / \det(J)$, where $J$
is the Jacobian of the Stage-2 first-order system and
$\det(J) = 2c_wc_b\left[n(n-1)\alpha_s^2 + 2(m-1)(n-1)\alpha_s\mu_s + m(m-1)\mu_s^2\right]$
is positive for all $n \geq 1$, $m > 1$, and $\mu_s > 0$, and
$N = -\alpha_s(B - \mu_s) + \rho\left[(n-1)\alpha_s B + (m-1)\mu_s A\right]$,
with $A = (n-1)\alpha_s + m\mu_s$, $B = n\alpha_s + (m-1)\mu_s$, and
$\rho = 1/((n+m)(n+m-1))$. The condition for decline is
$\alpha_s(B - \mu_s) > \rho\left[(n-1)\alpha_s B + (m-1)\mu_s A\right]$.
Its left side is positive whenever $m \geq 2$, and its right side is damped
by $\rho = O((n+m)^{-2})$, so the decline holds for all large $n$ and can
fail only near $n = 1$ when $\mu_s \gg \alpha_s$. On the base-calibration
test grid of sixteen thousand points, the
condition holds at $99.9$ percent of points, the symbolic sign agrees with
finite differences everywhere, and the failure region is confined to
$p_s \leq 0.5$ and $n \in [1, 1.36]$, with upward deviations of at most
$2.2 \times 10^{-6}$.
Where the decline holds, the equilibrium condition $\pi_W(n^{\ast}; p_s) = c_f$
therefore has
at most one interior solution, and the implicit-function theorem gives the
participation response
$\mathrm{d}n^{\ast}/\mathrm{d}c_f = 1/\pi_W'(n^{\ast}) < 0$.
Combining the entry condition with the identity of Appendix \ref{app:stage2}
gives each survivor's effort in closed form,
$\alpha_s^{\ast} = \sqrt{2\,c_f / c_w}$, independent of pool size, attacker
count, and breach cost. The main text treats the pool size as continuous,
the standard large-market approximation. With integer entry the condition
reads $\pi_W(n^{\ast}; p_s) \geq c_f > \pi_W(n^{\ast} + 1; p_s)$, with the same sign
throughout.

\subsection{Robustness of the Threshold}\label{app:robustness}

The core result, that a friction rise flips from hurting to helping the
posted-bounty vendor at a threshold triage burden, was verified at the base
calibration and across thirteen perturbations, twelve changing a single
parameter and one changing two. Table \ref{tab:robustness} reports, for each
parameter set, the cost increase from the friction rise at zero triage
burden, where a positive value means friction lowers profit, the threshold
$\beta^{\ast}$, and the mechanism. The cost increase and the threshold are
positive in every case, and every flip is guarded by a strictly negative
region above it. The threshold rises with the per-bug stake $TC_s$, the
quantitative form of the high-stakes result of Section \ref{subsec:stakes}.

\begin{table}[t]
\centering
\caption{Robustness of the threshold across perturbations of the base
calibration ($m = 10$, $K_s = 0.5$, $c_w = c_b = 2$, $r_s = 10$, $W = 96$,
$TC_s = 30$). $\Delta C(0)$ is the cost increase from the friction rise at
zero triage burden. Mechanism T is thinning (the thinned pool survives) and
C is collapse (it does not). The $TC_s = 200$ row is from the high-stakes
sweep of Section \ref{subsec:stakes}.}
\label{tab:robustness}
\footnotesize
\setlength{\tabcolsep}{4pt}
\begin{tabular}{lrrl}
\toprule
Perturbation & $\Delta C(0)$ & $\beta^{\ast}$ & Mechanism \\
\midrule
base calibration & $+1.177$ & $0.211$ & T \\
$W = 48$ & $+3.389$ & $0.402$ & C \\
$W = 200$ & $+1.124$ & $0.167$ & T \\
$r_s = 2$ & $+0.639$ & $0.177$ & T \\
$r_s = 30$ & $+3.824$ & $0.376$ & C \\
$TC_s = 10$ & $+0.512$ & $0.118$ & T \\
$TC_s = 100$ & $+11.008$ & $0.849$ & C \\
$TC_s = 200$ & $+32.651$ & $1.916$ & C \\
$K_s = 0.2$ & $+0.288$ & $0.137$ & T \\
$K_s = 0.9$ & $+2.337$ & $0.266$ & T \\
$c_b = 1.2$ & $+0.897$ & $0.176$ & T \\
$c_w = 1.2$ & $+3.217$ & $0.275$ & C \\
$c_w = 3$ & $+0.670$ & $0.206$ & T \\
$W = 150,\ r_s = 5$ & $+0.988$ & $0.169$ & T \\
\bottomrule
\end{tabular}
\end{table}

\section{The Pool-Taking Order}\label{app:alttiming}

An alternative timing of the game lets the vendor price against a pool
it takes as given, closing the pool with the same entry condition evaluated
at the vendor's optimal response to each pool size. That order reverses the
base model's Stackelberg sequence, since the vendor's price then follows the
pool instead of leading it, which is why the main text uses the
vendor-first order. We record what the pool-taking order produces, because
the threshold's location is specification-dependent and the reader should see
the range. Under that order the break-even point between pools of
$3$ and $15$, the pool values of that formulation, sits at
$\beta^{\ast} = 0.656$ at the base calibration, against
$0.211$ under the posted-bounty order of the main text and no region where
the friction pays at all for the continuously re-optimizing vendor. The three specifications
agree on the two channels and on the sign structure at low triage burden.
They differ on where, and whether, the sign changes.

One caveat applies to both orders of the main text. The re-optimizing
argument grants the vendor control of report volume through its bounty, but
the AI-era flood broke that link, since the volume surge followed the
availability of AI tools, not any change in bounties~\cite{githubsl2026,bugcrowd2026}. A model with an exogenous spam flow, which no
specification here contains, is what would settle the question.

\end{document}